# Darwin's Neural Network: AI-based Strategies for Rapid and Scalable Cell and Coronavirus Screening

Sang Won Lee[1], Yueh-Ting Chiu[1], Philip Brudnicki[1], Audrey M. Bischoff[1], Angus Jelinek[1], Jenny Zijun Wang[1], Danielle R. Bogdanowicz[1], Andrew F. Laine[1,2], Jia Guo[3], and Helen H. Lu[1*]

[1] Department of Biomedical Engineering, Columbia University, New York, NY, USA.
*e-mail: hhlu@columbia.edu
[2] Department of Radiology, Columbia University, New York, NY, USA.
[3] Department of Psychiatry, Columbia University Medical Center, New York, NY, USA.

**Abstract.** Recent advances in the interdisciplinary scientific field of machine perception, computer vision, and biomedical engineering underpin a collection of machine learning algorithms with a remarkable ability to decipher the contents of microscope and nanoscope images. Machine learning algorithms are transforming the interpretation and analysis of microscope and nanoscope imaging data through use in conjunction with biological imaging modalities. These advances are enabling researchers to carry out real-time experiments that were previously thought to be computationally impossible. Here we adapt the theory of survival of the fittest in the field of computer vision and machine perception to introduce a new framework of multi-class instance segmentation deep learning, Darwin's Neural Network (DNN), to carry out morphometric analysis and classification of COVID19 and MERS-CoV collected *in vivo* and of multiple mammalian cell types *in vitro*.

## 1. Introduction

Coronavirus disease-19 (COVID-19) is an emerging acute respiratory infectious disease that has demonstrated highly pathogenic capabilities, spreading through populations globally primarily through droplet transmission. Although the virus is expected to be of zoonotic origin in the seafood markets of Wuhan, China, global human-to-human transmission has prompted the emergence of over 15 million COVID-19 cases worldwide and over five hundred thousand deaths. COVID-19 is the seventh member of the family of coronaviruses to widely cause infection in humans. [1] The clinical spectrum of coronavirus ranges from asymptomatic forms to conditions characterized by respiratory failure to septic shock. [2] The first known widespread infection caused by a coronavirus began in 2002, with the emergence of severe acute respiratory syndrome coronavirus (SARS-CoV) in China, that resulted in 8,098 infections and 774 deaths among 29 countries. [3] A second outbreak followed in 2012, beginning in Saudi Arabia, with the spread of the Middle East Respiratory Syndrome coronavirus (MERS-CoV) that infected 2,458 people and resulted in 848 deaths in 27 countries. [4] COVID-19 is the third coronavirus outbreak of the 21st century, and it is already more deadly than the previous outbreaks. However, the ability to combat this emerging infectious disease has been limited by the slow turnaround time in the development of new therapeutics, the inability to quickly diagnose patients, and the limited knowledge of the virus' pathogenesis.

The pathophysiology and virulence mechanisms of coronaviruses have been shown to be mediated through the virion morphological structure and surface structural proteins. [5] Coronaviruses have distinct morphological features that make them easily distinguishable under a high-powered microscope. Typical coronavirus virions are spherical, 125 nm in diameter, and have club-shaped projections emerging from their surfaces. [6] MERS-CoV is comprised of four major surface proteins that aid in viral infiltration of cells: envelope protein (E), spike glycoprotein (S), nucleocapsid protein (N), and membrane protein (M). [7] Each protein has an integral function in virus transmission within a host. For instance, the S protein that comprises the spikes on the surface mediates virus entrance via binding and fusion to host cells, as it contains a receptor domain that binds to the host cell receptors. [7] Currently, extensive knowledge regarding the structure and morphology of the COVID-19 virus limits the understanding of its pathogenesis and virulence. However, as the spikes are a unique characteristic of coronaviruses, including COVID-19, common approaches in therapeutic development to neutralize their viral infections involve inhibition of surface protein capabilities, such as blocking the interaction between the S protein and its host receptor. [8] Given the previous success with using animal models to study *in vivo* the ability of antibodies and other therapeutics to limit viral replication as well as the pathology of MERS-CoV, [9, 10] similar therapeutic approaches can be developed to analyze morphological effects on COVID-19 *in vivo* in response to similar therapies. This paper intends to implement novel machine learning methods to analyze the *in vivo* morphology of COVID-19, comparing it to the better-known MERS-CoV. Evaluation of the emerging viral strain, collected *in vivo*, under the microscope and comparing it to the existing MERS-CoV morphology will potentially allow for a better understanding of the virus' pathogenesis. We have correctly classified different types of coronaviruses using a deep learning multi-class instance segmentation network as well as analyze their morphological properties.

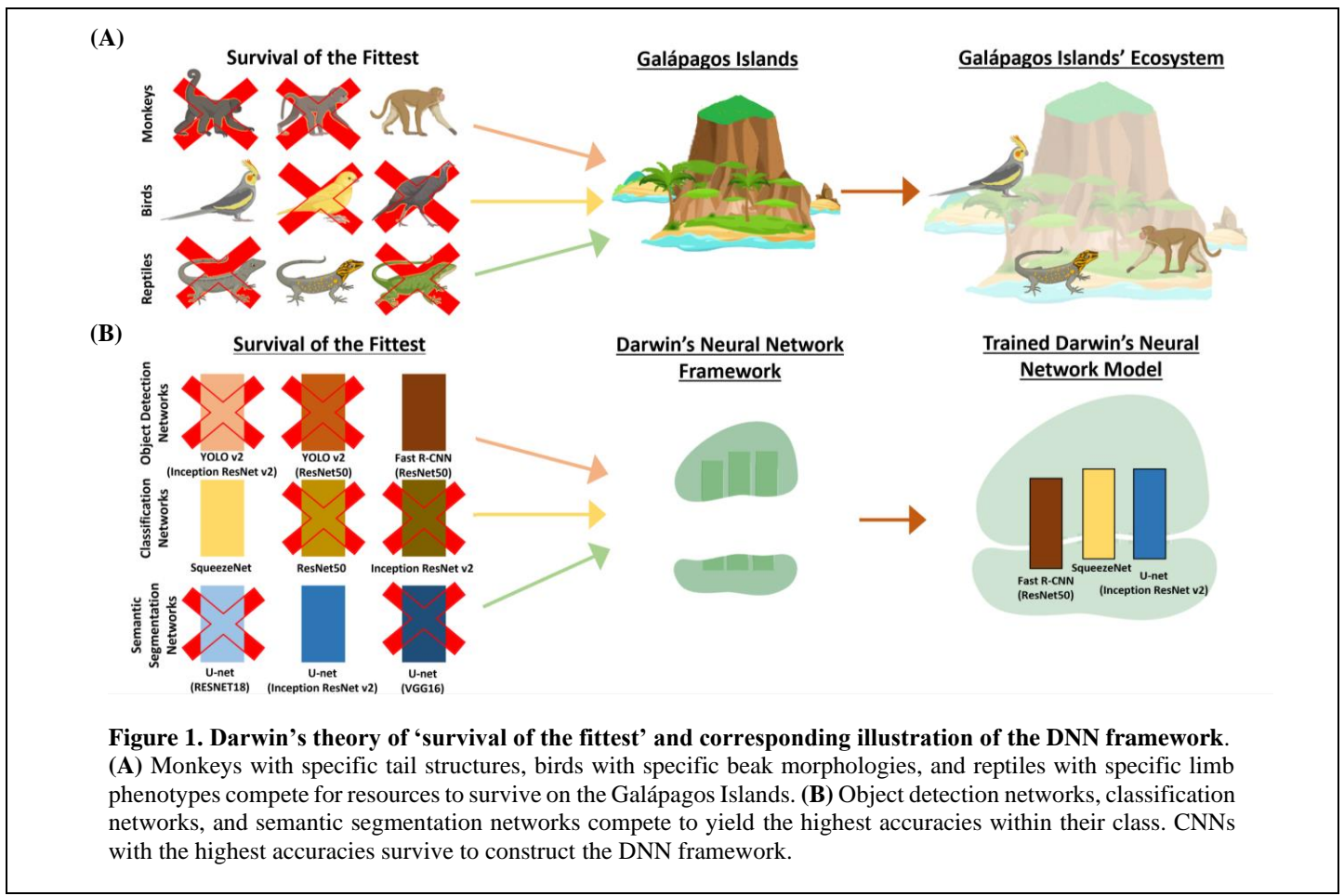

**Figure 1. Darwin's theory of 'survival of the fittest' and corresponding illustration of the DNN framework**. **(A)** Monkeys with specific tail structures, birds with specific beak morphologies, and reptiles with specific limb phenotypes compete for resources to survive on the Galápagos Islands. **(B)** Object detection networks, classification networks, and semantic segmentation networks compete to yield the highest accuracies within their class. CNNs with the highest accuracies survive to construct the DNN framework.

Advances in the field of deep learning are allowing previously thought impossible research to be conducted. Every year, new convolutional neural networks (CNNs) yield higher accuracies for their tasks with higher GPU efficiency. Typical tasks for CNNs include object tracking, image classification, and semantic segmentation. Object tracking allows following of an entity, such as tracing the migration of a cell; image classification is used to predict a label for an object, such as determining whether a cell is a type I or type II macrophage; and semantic segmentation identifies parts of an image that correspond to distinct objects, such as identifying pixel locations of a nucleus in a mammalian cell. However, optimal neural networks for a certain task change every year due to the invention of newer and more powerful CNNs. In classification neural networks, for example, AlexNet [11] is rarely used except for educational purposes after the publication of superior neural networks such as GoogLeNet [12], followed by VGG-16 [13], Inception-ResNet-v2 [14], and NASNet-Large. [15] As for object detection networks, R-CNN [16] was triumphed by Fast R-CNN [17], which was surpassed by Faster R-CNN. [18] YOLO [19] network was surpassed by YOLO2. [20] Finally, for semantic segmentation neural networks, the original U-Net [21] evolved to yield higher accuracy by adapting newer neural networks like VGG-16 into its architecture. Thus, the capabilities of classification, object detection, and segmentation networks are continually adapting and succeeding their predecessors for different tasks.

Here we propose a new framework for multi-class instance segmentation that utilizes three independent CNNs: object detection network, classification network, and semantic segmentation network. As state-of-the-art networks emerge in each field, pre-existing CNNs and the new CNN compete to yield the highest accuracies for the task, and only the CNNs with superior accuracies survive to form a framework. One can also choose to cull the neural networks and replace them to yield a higher multi-class instant segmentation accuracy. Using the combined CNNs, the framework can automatically comb through the existing CNNs and select the combination with superior reliability and accuracy. We call this conglomerate Darwin's Neural Network, as the "fittest" or most accurate results yielding CNNs survive and are replaced over time. A graphic illustration of the DNN framework is shown in **Figure 1**. This network can be implemented to compare morphometric parameters of MERS-CoV and COVID19 virus particles using transmission electron micrographs (TEM). Specifically, we wanted to use these micrographs to investigate structural and morphological changes in these virus particles *in vivo*.

Advances in microscopy have enabled researchers to access spatial and temporal variations inherent in biological systems. Progress in the field of optics has resulted in microscopes capable of imaging over a range of spatial scales, from single cells to organisms in its entirety. In concurrence with these technological advances, there has been an overwhelming increase in demand in the biosciences for automatic and precise image analysis. Here we also implement DNN to establish an automated method for cell morphometric analysis and cell-type classification utilizing only brightfield images taken on a benchtop microscope directly from cell-culture wells. Despite the low resolution of the images obtained and significant well-to-well heterogeneity, our team was able to demonstrate precise morphometric analysis and high classification accuracy of novel data.

There exist three major components in DNN: object detection network, classification network, and semantic segmentation network. A graphic illustration of DNN workflow is shown in **Figure 2**. First, multiple object detection networks compete, and the winner only finds locations of morphologically similar objects of interest and crop them out automatically to feed them to more GPU exhaustive and accurate classification networks. At this stage, the classification task is not carried out between the morphologically similar objects but leaves the heavy lifting for a more apt classification network. The objects' location information is saved for the reconstruction of images at a later stage of DNN framework. Then, multiple classification networks compete, and the winner takes in the cut-out images as inputs and carries out classification task of morphologically very similar objects; for example, COVID19 virus particles to MERS virus particles, or macrophage type I to macrophage type II. Then, the cropped images and their classes are passed onto the segmentation network for semantic segmentation according to their class for instance segmentation and accurate morphological analysis. These steps result in instance segmentation instead of semantic segmentation. This instance segmentation network can be used for tasks, such as single-cell morphological analysis in clusters or colonies of cells and proves to be more accurate than any algorithm for post semantic segmentation to singularize objects in binary clusters. Each segmentation result can be colored differently and labeled accordingly to their classes. The results are superimposed on top of the original input image for object detection to achieve a multi-class instance segmentation network framework.

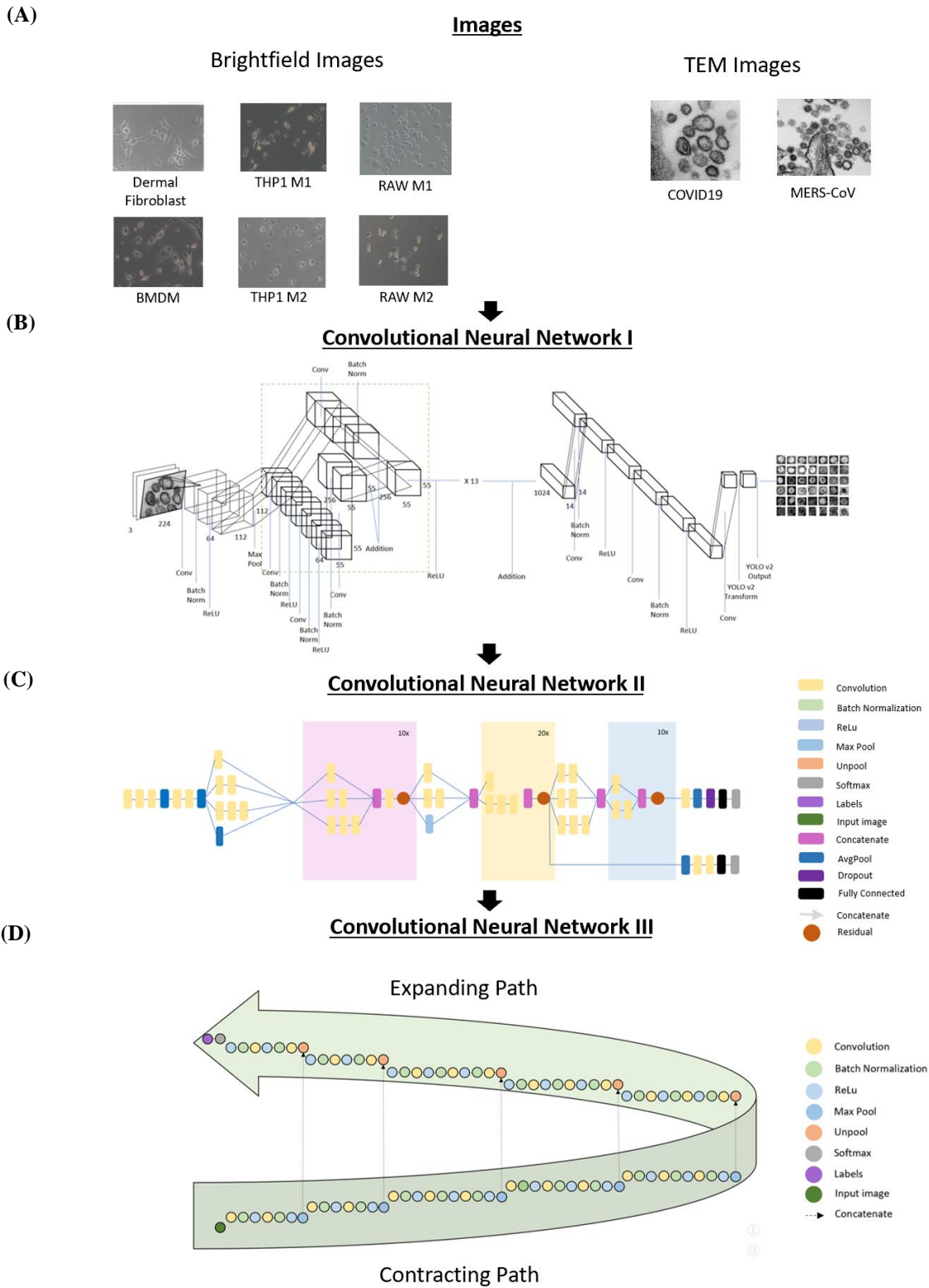

**Figure 2. Illustration of DNN workflow. (A)** Brightfield images and TEM images are the inputs to the DNN. **(B)** The input images are fed into CNN I, an object detection network, to acquire singularized images of cells and viruses. The model architecture shown here is YOLO v2 with ResNet50 backbone. **(C)** The singularized images of cells and viruses are fed into CNN II, a classification network, for identification of their innate types. The model architecture shown here is Inception-ResNet-v2. **(D)** The identified cells are then fed into CNN III, a semantic segmentation network, where results of morphometric data are produced. The model architecture shown here is U-Net with a VGG16 backbone.

## 2. Materials and Methods

2.1 Data Acquisition and Preparation

2.1.1 THP-1 Cell Culture

Human THP-1 cells (ATCC, TIB-202) were commercially obtained from ATCC. Cell cultures were maintained in suspension in non-tissue culture treated flasks (Nunc) with a surface area of 25 cm$^2$. The culture media was refreshed every 2 days and was comprised of Roswell Park Memorial Institute (RPMI) 1640 Medium supplemented with 10% fetal bovine serum (Atlanta Biologics), 1% penicillin-streptomycin (Sigma-Aldrich) and 0.05 mM 2-mercaptoethanol (Sigma-Aldrich). Cells were initially suspended at 200,000 cells / mL of media and passaged upon reaching a density of 1.0 million cells / mL.

For this study, passage two cells were collected and resuspended at 100,000 cells / mL in fresh fully supplemented RPMI (F/S RPMI) media containing 100nM phorbol 12-myristate 13-acetate (PMA) to induce differentiation. One mL of this cell suspension was then added to each well of a 12-well plate (BD Falcon) and cultured at 37°C. After 72 hours, PMA containing medium (PMA + medium) was removed, and adherent cells were rinsed with PBS, and the medium was replaced with F/S RPMI-1640 medium. Cells were allowed to rest (for M0) or were subjected to polarization media for 72 hours before assessment. M1 polarization medium contains F/S RPMI + 20 ng/mL interferon-γ (IFN-γ, Humanzyme) and 1 µg/mL lipopolysaccharide (LPS, Sigma-Aldrich). M2 polarization medium contained F/S RPMI + 20 ng/mL interleukin-4 (IL-4, Humanzyme). After 72 hours of rest or polarization, cells were washed with PBS 1x and cultured in F/S RPMI media. All groups were prepared with a batch size n=10. After polarization, cells were imaged (n=10) using brightfield microscopy with a phase-contrast filter. Images for machine learning analysis were captured at 32x, and each frame contained approximately 20 cells.

2.1.2 RAW 264.7 Cell Culture

RAW 264.7 cells (ATCC, TIB-71) were commercially obtained from ATCC. Cell cultures were maintained in 100 mm non-tissue culture treated plates (Fisher). The culture media was refreshed every two days and was comprised of Dulbecco's Modified Eagle's Medium (DMEM) supplemented with 10% fetal bovine serum (Atlanta Biologics), and 1% penicillin-streptomycin (Sigma-Aldrich). Cells were initially suspended at 200,000 cells per dish and passaged upon reaching a 75% confluency.

For this study, passage two cells were collected and seeded into 48 well plates at a density of 25,000 cells / cm$^2$. After 18 hours to allow cell attachment, cells were either allowed to rest or treated with polarization media for 48 hours. For the M0 phenotype, the cells were allowed to culture in F/S DMEM media (as described above). For M1 polarization, F/S DMEM containing 10 ng/mL LPS (Sigma-Aldrich) and 20 ng/mL IFN-γ (Humanzyme) was used. For M2 polarization, F/S DMEM containing 20 ng/mL IL-4 (Humanzyme) was used. After 48 hours, cells were rinsed with PBS and cultured in F/S DMEM just prior to imaging. Cells were imaged (n=10) using brightfield microscopy with a phase-contrast filter. Images for machine learning analysis were captured at 32x, and each frame contained approximately 20 cells.

2.1.3 Dermal Fibroblast Isolation and Cell Culture

Following published protocols [22], bovine dermal fibroblasts (DF) were harvested from bovine skins. Briefly, neonatal (1-7 days old) bovine skins were obtained from a local abattoir (n=2, tissues pooled; Green Village Packing Company). Before harvest, skins were sterilized by soaking in soapy water for 40 min, followed by 70% ethanol for 20 min, after which the surrounding fur was removed with a sharp scalpel. Approximately 1 cm$^2$ skin fragments were collected aseptically in a sterile environment. The dermis was separated from the epidermis by gently scraping dermis with a scalpel. The dermis was digested for 30 min at 37 °C with collagenase II (1.2% w/w; Worthington Biochemical) in Dulbecco's Modified Eagle's Medium (DMEM, Cellgro-Mediatech) supplemented with 10% fetal bovine serum (FBS), 2% penicillin/streptomycin (P/S), 0.2% Amphotericin B (Amp-B), and 0.2% Gentamicin (G/S). The mixture was then filtered (30 µm, Spectrum Labs), and the isolated cells were collected by centrifugation and plated. After 14 days from the beginning of cell isolation, the cells were re-plated at a density of 5 × 10$^5$ cells/cm$^2$ on tissue culture plates. Images for machine learning analysis were captured at 32x, and each frame contained approximately 20 cells.

2.1.4 Bone Marrow Derived Macrophage Isolation and Cell Culture

Bone marrow derived macrophages (BMDM) were harvested from the murine tibia. 4-8 weeks old Macrophage Fas-induced Apoptosis mice were purchased (n=2; The Jackson Laboratory). Prior to harvest, mice were sterilized by soaking in soapy water for 40 min, followed by 70% ethanol for 20 min, after which the tibiofemoral joints were removed. The surrounding subcutaneous fascia and muscle were removed aseptically in a sterile environment. Tibial tuberosity and medial malleolus were removed from tibia. Bone marrow cells were flushed out by forcing Roswell Park Memorial Institute (RPMI; Thermo Fisher Scientific) containing 5% FBS through the central bone marrow canal using a 10 ml syringe. The collected bone marrow tissue was then filtered (70 µm, Spectrum Labs), and the isolated cells were collected by centrifugation. The cells were mixed with 1 mL of ammonium chloride (ACK) lysis solution and were promptly washed with 1 mL of RPMI media containing 5% FBS. The isolated cells were then collected by centrifugation and plated at a density of $1 \times 10^7$ cells/cm$^2$ on tissue culture plates (non-treated 100Π Petri Dish). Images for machine learning analysis were captured at 32x, and each frame contained approximately 20 cells.

2.1.5 Virus Isolation

Transmission electron micrographs (TEM) of COVID19 and MERS-CoV virus particles isolated from patients, were obtained from the open database published by National Institute of Allergy and Infectious Diseases' (NIAID) Rocky Mountains Laboratories (RML).

2.2 Convolutional Neural Networks

Three different types of CNNs were considered for the DNN deep learning algorithm: object detection networks, classification networks, and semantic segmentation networks.

2.2.1 Convolutional Neural Network I

For CNN I, YOLO v2 and Faster-R-CNN were used with ResNet50 and Inception-ResNet-v2 backbones. These four CNNs were transfer-learned and tasked to isolate individual cells in brightfield microscope images, as seen in **Figure 3(A).** The networks were trained to only isolate the cells, which showed the complete morphology. The networks were trained not to pick up overlapped cells since missing part of the data can skew later morphometric analysis. Another set of CNNs was transfer-learned and tasked to isolate COVID19 and MERS-CoV virus particles in transmission electron micrographs. Again, the networks were trained to only isolate the viruses, which showed the complete morphology.

The output coordinates were modified to superimpose boxes onto the original image and crop each object detection result, as shown in **Figure 3(B).** Input images were rotated with mirrored corners to increase the size of the training set. A total of 217 TEM micrographs of COVID19 and MERS-CoV virus particles were used. 130 TEM images were used for the training set, 65 images were used for the validation set, and 22 images were used for the test set (6:3:1). For the cells, 540 brightfield images were used for the training set, 270 images were used as a validation set, and 90 images were used for the test set. The training sets were carried out until absolute minima were observed for the loss function. Other parameters, such as kernel, stride, max pooling sizes were unadjusted to retain the advantages of original CNNs and maximize the benefit of transfer learning. The network which produced the highest precision over recalls was chosen to be integrated into the DNN. The chosen CNN was used to crop individual cells from TEM and brightfield feeds. The resulting cropped images were used to train Convolutional Neural Network II (CNN II). The resulting images were further processed to greyscale images, and the histograms of images were equalized to reduce bias.

2.2.2 Convolutional Neural Network II

For CNN II, Inception-ResNet-v2, Squeezenet[29], Mobilenet-v2[30], Resnet18[31], ResNet 50[32], ResNet101[33], Inception v3[34], VGG16, VGG19[35], DenseNet201[36], Xception[37], AlexNet, and GoogLeNet were used to compete with each other. The individually cropped cells and viruses from CNN I were manually divided into respective classes to create the training sets, as seen in **Figure 3(C).** Again, the training images were rotated and mirrored to increase the training set. A total of 1680 brightfield images of cells was used for CNN II. 1008 training images, 504 validation images, and 168 test set images (6:3:1). A total of 360 images of virus particles was used: 216 training images, 108 validation images, and 36 test set images. The CNN, which yielded the highest accuracy, was integrated into the DNN to carry out the task. To visualize the progress and focus of the CNN, activation maps were derived from the last rectifier linear unit. Activation maps were created for iteration 1, iteration 5, and iteration 700 for viruses. Visualization maps of completed CNN were created for cells and their corresponding classes.

2.2.3 Convolutional Neural Network III

U-Nets with ResNet18, ResNet50, VGG16, and Inception-ResNet-v2 backbones competed with each other for placement in CNN I. Corresponding masks were manually created for cells and viruses according to their class as seen in **Figure 3(D).** A total of 360 TEM images and their corresponding 360 masks of virus particles were used for CNN II, utilizing 216 images for the training set, 108 images for the validation set, and 36 images for the test set (6:3:1). A total of 1680 brightfield images and their corresponding 1680 masks of cells were used for CNN II. 1008 training images, 504 validation images, and 168 test set images (6:3:1). The network with the highest global accuracy, as determined by the ratio of correctly classified pixels to the total number of pixels, regardless of class, was integrated into DNN. The resulting binary output images were passed down for further morphometric analysis.

2.3 Mask-R-CNN

Facebook's Mask-R-CNN [23] with Microsoft's ResNet101[24] backbone was used to compare instance segmentation results. Jaccard Similarity Coefficient was used to evaluate both Mask-R-CNN and DNN.

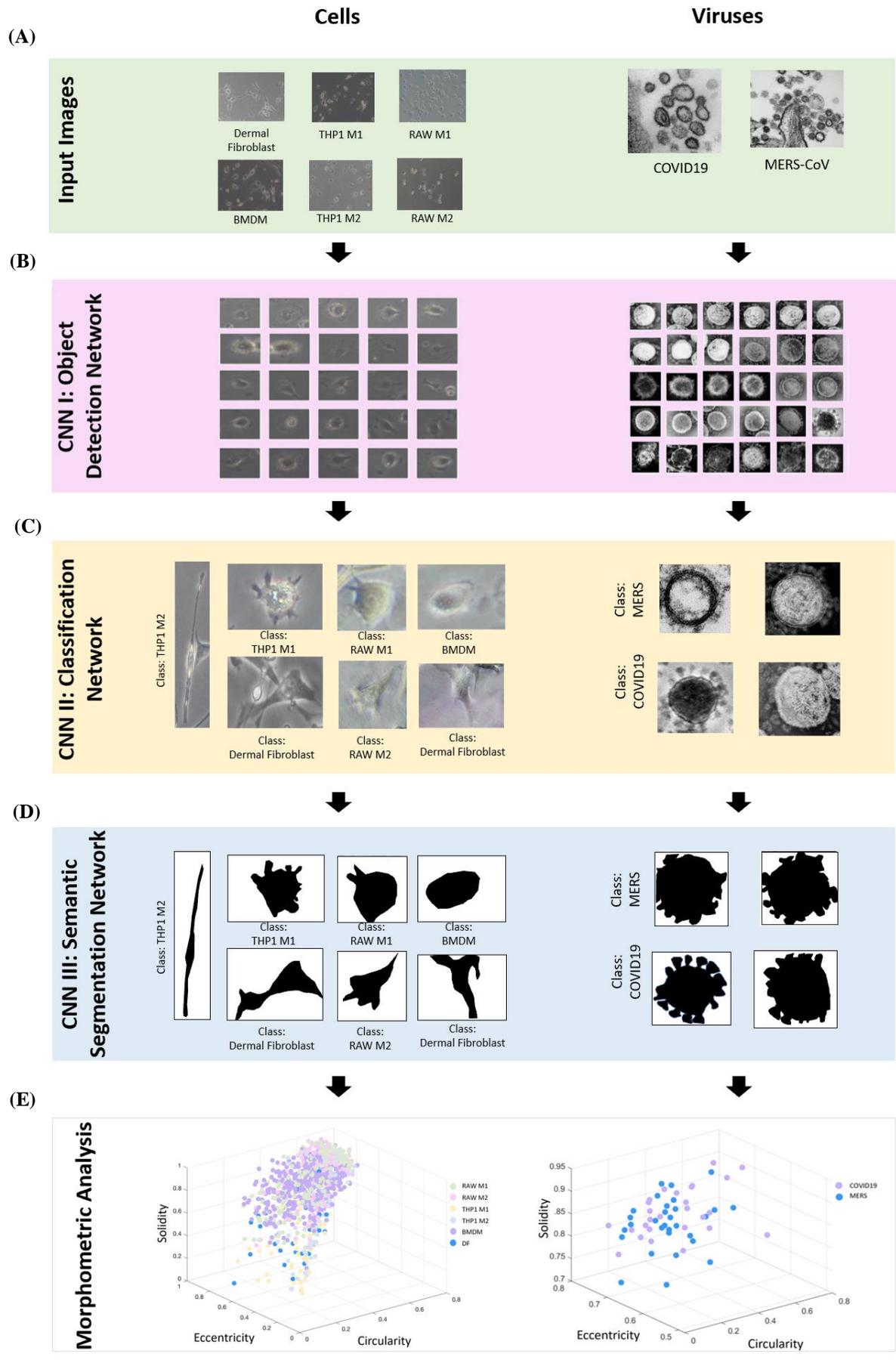

**Figure 3. Inner workings of DNN. (A)** Input images that are prepared by brightfield microscope or TEM. These input images are fed into CNN I. **(B)** Individualized cells and viruses by CNN I, object detection network. These are also outputs of CNN I and inputs to CNN II. **(C)** Classified cells and viruses using CNN II, classification network. These individually classified and singularized cells and viruses are outputs of CNN II and inputs of CNN III. **(D)** Semantic segmented cells and viruses using CNN III, semantic segmentation network. These binary images are outputs of CNN III and input to further morphometric analysis. **(E)** Morphometric data of cells and viruses are derived from the binary inputs, which are the outputs of CNN III. The following morphometric parameters are calculated: area, eccentricity, circularity, and solidity.

2.4 Post Machine Learning Processing.
    Binary image outputs of CNN III were further processed using the regionprops function in MATLAB® [25] to calculate morphometric parameters of the virus.

2.5 Morphometric Analysis
    The following formulas were used for morphometric analysis:
Circularity: $(4\pi \times Area)/convex\ perimeter^2$
Solidity: The proportion of the pixels in the convex hull that are also in the object. [26]
Eccentricity: The eccentricity is the ratio of the distance between the foci of the ellipse and its major axis length. The value is between 0 and 1. [26]

2.6 Statistical Analysis
    Results are presented as the mean ± standard deviation. The Tukey–Kramer posthoc-test was used for all morphometric pairwise comparisons, and statistical significance was attained at $p < 0.05$.

## 3. Results

### 3.1 Test Set Accuracies
3.1.1 CNN I
    The results for CNN I are shown in **Figure 4(A, B).** Precision was one over all recalls for YOLOv2 with ResNet50 [24] and Inception-ResNet-v2 for both cells and viruses. Faster-R-CNN also yielded identical precision over all recalls for ResNet 50 and Inception-ResNet-v2 backbones for both cells and viruses. Since all architectures yielded perfect precision, Faster-R-CNN with ResNet50 backbone was chosen to be integrated into the place of CNN I in DNN for cells. For the viruses, YOLO v2 with ResNet50 backbone was chosen to be in place of CNN I in the DNN framework.

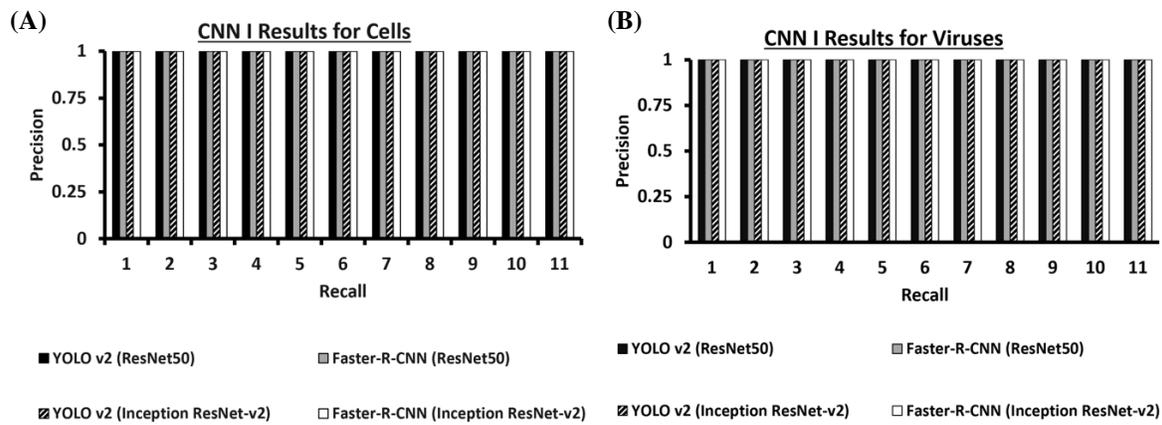
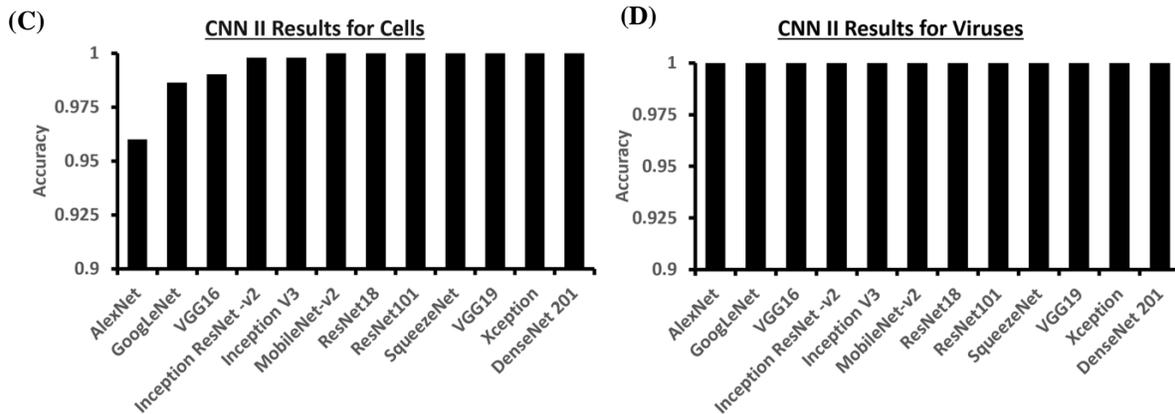
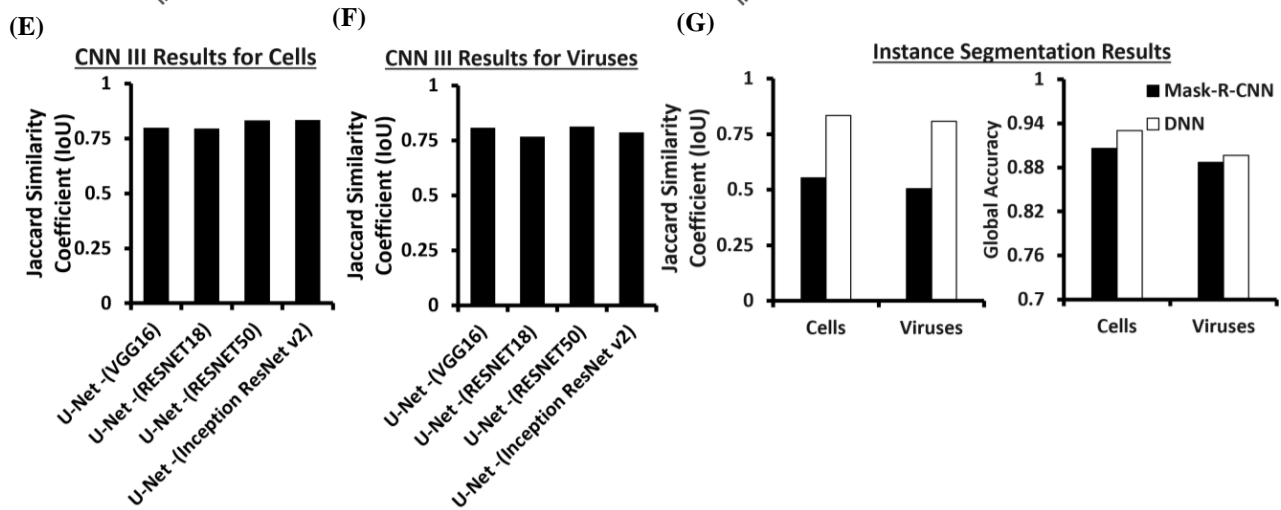

**Figure 4. Resulting precisions, accuracies, and Jaccard similarity coefficients for CNN I, II, and III. (A, B)** Precision over recall results of CNN I, object detection networks, for different neural architectures for cells and viruses, respectively. **(C, D)** Accuracy results of CNN II, classification networks, for different neural architectures for cells and viruses, respectively. **(E, F)** Jaccard similarity coefficients of CNN III, semantic segmentation networks, for different neural architectures for cells and viruses, respectively. **(G)** Jaccard similarity coefficients and global accuracy comparison for Mask-R-CNN and DNN.

3.1.2 CNN II

The results for CNN II are shown in **Figure 4(C, D).** For CNN II cell classification, AlexNet yielded the lowest test set accuracy of 0.96 for the test set. GoogLeNet yielded the second-lowest accuracy of 0.985. Rest of the networks, DenseNet 201, Inception-ResNet-v2, Inception v3, MobileNet v2, ResNet18, ResNet101, SqueezeNet, VGG19, Xception, yielded an accuracy of 1 for all test sets. For virus classification, all networks yielded an accuracy of 1 for the test sets. For visualization of progression and focus of the CNN, activation maps were created for iteration 1, iteration 5, and iteration 700 for viruses, as seen in **Figure 5(C).** Visualization maps of completed CNN were created for cells and their corresponding classes, as seen in **Figure 5(A, B).** From the neural architectures, which yielded an accuracy of 1, SqueezeNet was chosen to be integrated into the place of CNN II in the DNN framework for viruses. For cells, Inception-ResNet-v2 was chosen to be integrated into the DNN framework.

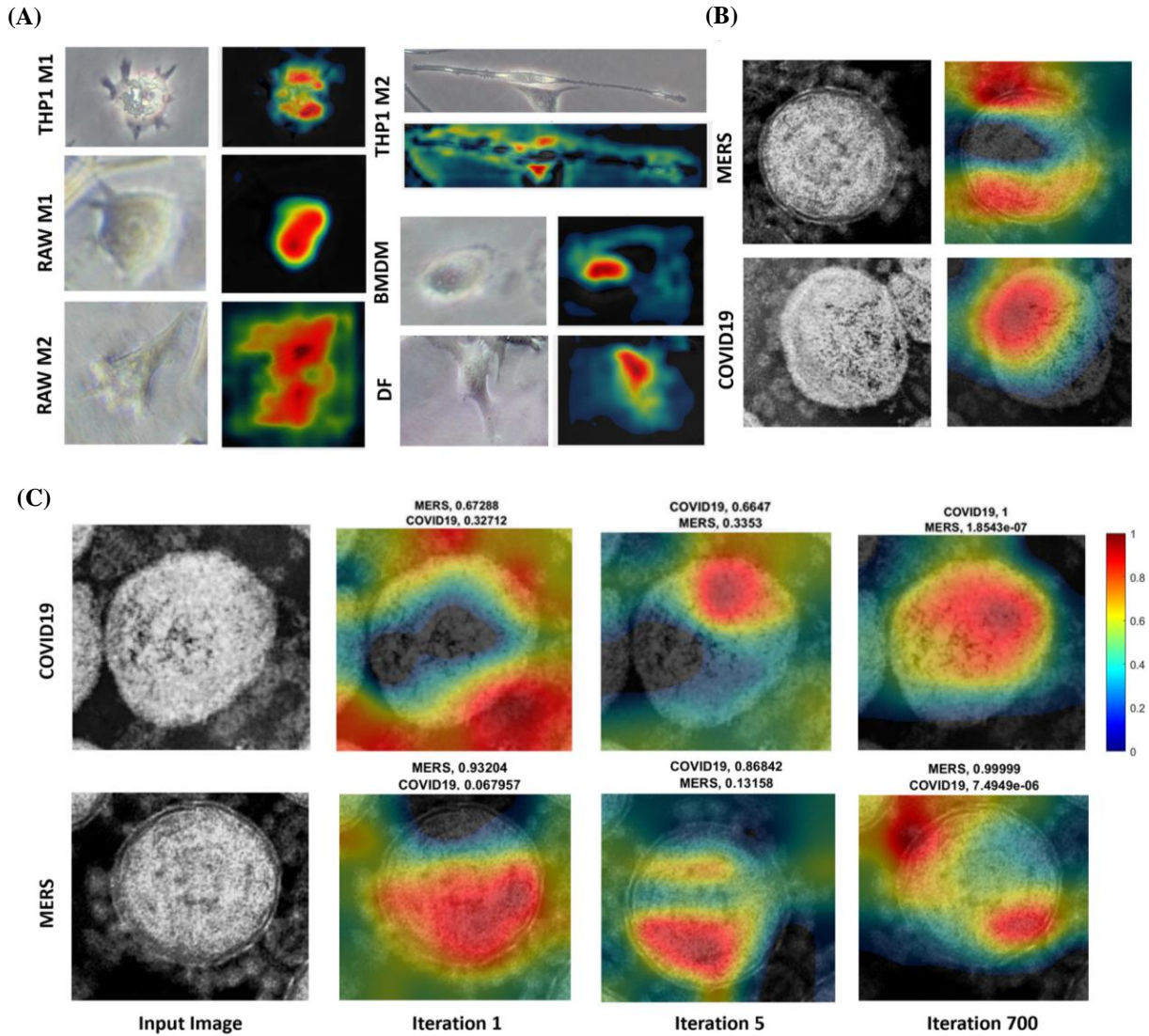

**Figure 5. Activation maps and their progression through training. (A, B)** Activation maps of each cell and virus types derived from transfer learned SqueezeNet in CNN II. Predictions are based on the cell and virus morphologies rather than the background. **(C)** Progression of activation maps derived from SqueezeNet according to their training iteration in CNN II. Bases for predictions are migrating from background to viruses as training progresses.

### 3.1.3 CNN III

The results for CNN III are shown in **Figure 4(E, F).** For CNN III cell semantic segmentation, U-Net with ResNet18 backbone yielded the lowest Jaccard similarity coefficient of 0.7942. U-Net with VGG16 yielded the second-lowest Jaccard similarity coefficient of 0.7984, and U-Net with ResNet50 yielded global accuracy of 0.8324. U-Net with Inception ResNet v2 backbone yielded the highest global accuracy of 0.8346, as seen in **Figure 4(E)**; therefore, Inception-ResNet-v2 was integrated in the place of CNN II for DNN for cells. For virus semantic segmentation, U-Net with ResNet18 backbone yielded the lowest Jaccard similarity coefficient of 0.7681 as seen in **Figure 4(F).** U-Net with VGG16 yielded a Jaccard similarity coefficient of 0.8083, and U-Net with ResNet50 yielded the highest Jaccard similarity coefficient of 0.8245. U-Net with Inception ResNet v2 backbone yielded a Jaccard similarity coefficient of 0.787; therefore, ResNet50 was chosen to be integrated into DNN for viruses.

### 3.1.4 DNN

As a result of the competition between the networks, DNN framework for cells consisted of Faster-R-CNN with ResNet50 backbone for CNN I, Inception-ResNet-v2 for CNN II, and U-Net with Inception-ResNet-v2 backbone. DNN framework for viruses consisted of YOLO v2 with ResNet50 backbone for CNN I, SqueezeNet for CNN II, and U-Net with ResNet50 backbone.

### 3.1.5 Mask-R-CNN

For overall instance segmentation results, DNN produced both superior global accuracy and Jaccard Similarity Coefficient for cells and viruses. For Mask-R-CNN, the global accuracies were 0.9059 and 0.8871 for cells and viruses, respectively, as seen in **Figure 4(G).** For DNN, the global accuracies were 0.9301 and 0.8964 for cells and viruses, respectively. For Mask-R-CNN, the Jaccard similarity coefficients were 0.5537 and 0.5038 for cells and viruses, respectively, as seen in **Figure 4(G).** For DNN, the Jaccard Similarity Coefficients were 0.8346 and 0.8083 for cells and viruses, respectively.

### 3.2 Morphometric Analysis

All results of the cellular and viral morphometric analyses are shown in **Figure 6**. All cells were plotted in a 3D graph according to their circularity, eccentricity, and solidity in **Figure 3(E).** Viruses were plotted in a 3D graph according to their circularity, eccentricity, and solidity in **Figure 3(E).** For cells, ground truth for area, circularity, eccentricity, and solidity were calculated by hand and compared with DNN output in **Figure 6(A1-A4).** For viruses, ground truths for area, circularity, eccentricity, and solidity were also calculated by hand and compared with DNN output in **Figure 6(B1-B4).** Statistical significances between the virus types, in terms of area and morphology, in the DNN output data are shown in **Figure 6(C2)** (n=33, ^p<0.05 between groups). Statistical significances between the virus types, in terms of area and morphology, in the ground truth data are shown in **Figure 6(C1)** (n=33, ^p<0.05 between groups). Statistical significances between THP1 M1 and THP1 M2, in terms of area and morphology, in the ground truth data are shown in **Figure 6(D1)** (n=51 and n=60, respectively. ^p<0.05 between groups). Statistical significances between THP1 M1 and THP1 M2, in terms of area and morphology, in the DNN output data are shown in **Figure 6(D1)** (n=51 and n=60, respectively. ^p<0.05 between groups). Statistical significances between the cell types, in terms of area and morphology, in the DNN output data are shown in **Table 1.**

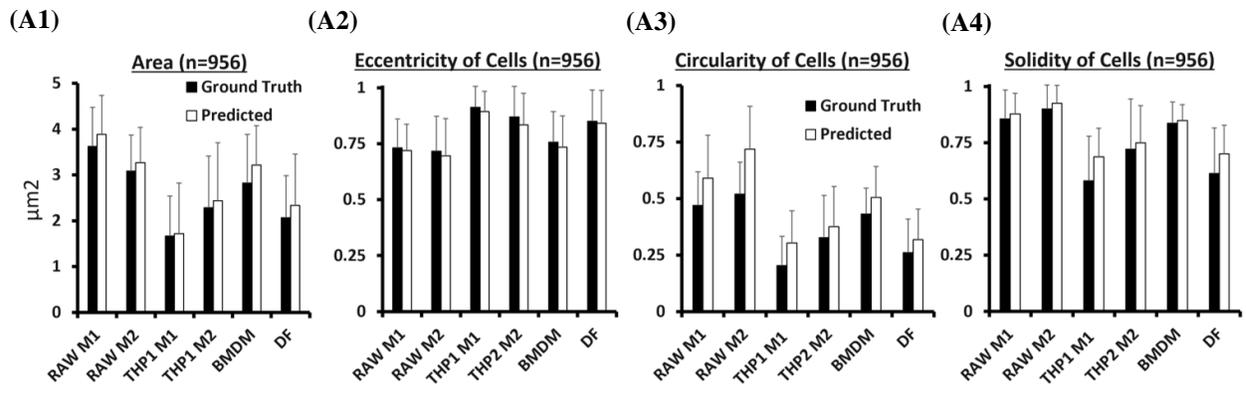
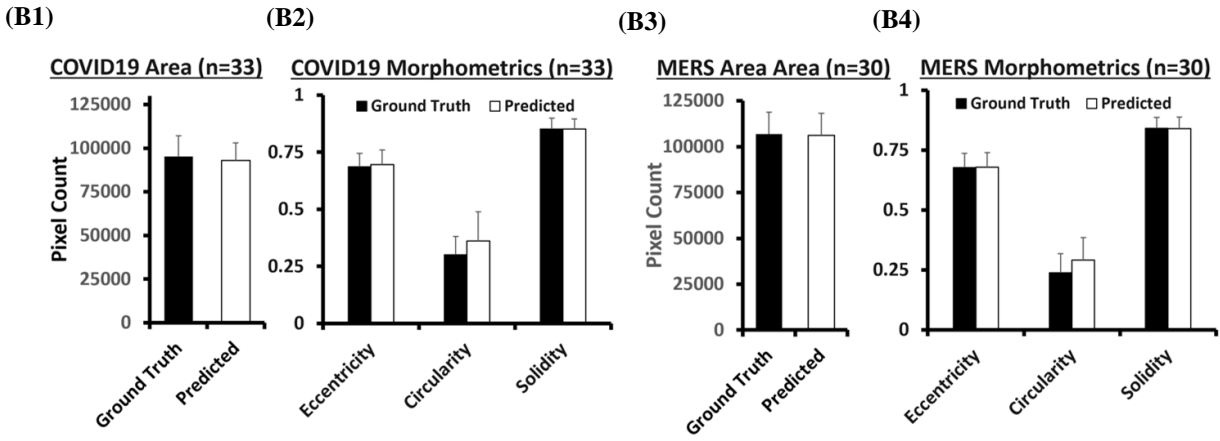
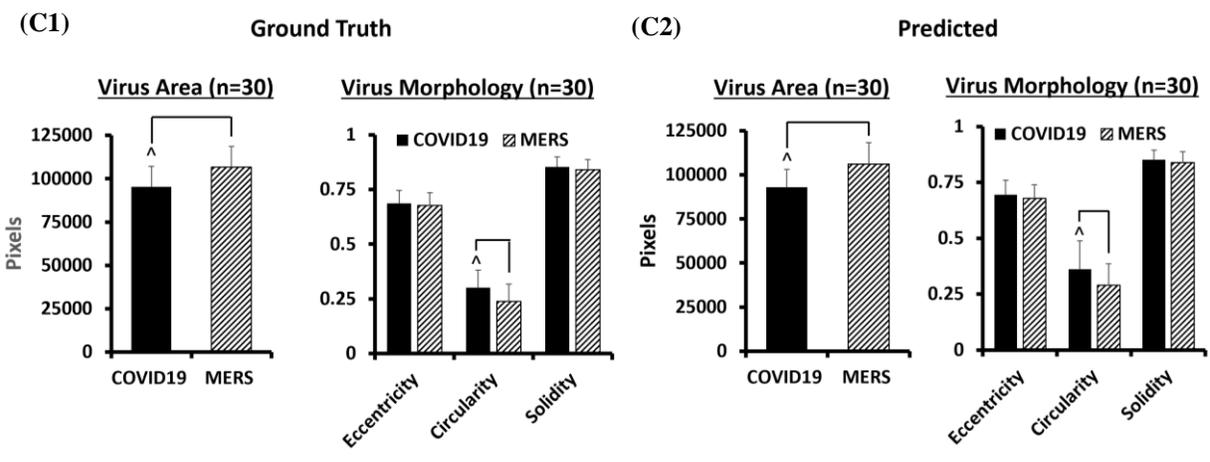
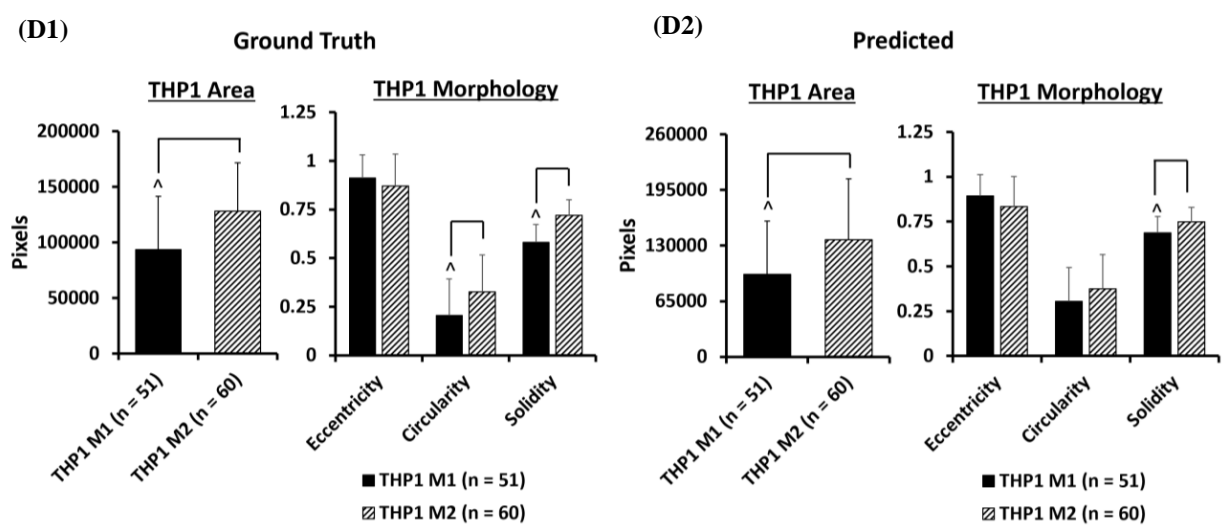

**Figure 6. Morphometric Analysis of cells and viruses (A1)** Cell areas calculated by hand compared to outputs of DNN. **(A2)** Eccentricity of cells calculated by hand compared to DNN outputs. **(A3)** Circularity of cells calculated by hand compared to DNN outputs. **(A4)** Solidity of cells calculated by hand compared to DNN outputs. **(B1)** COVID19 area calculated by hand compared to outputs of DNN. **(B2)** Eccentricity, circularity, solidity of COVID19 calculated by hand compared to DNN outputs. **(B3)** MERS area calculated by hand compared to outputs of DNN. **(B4)** Eccentricity, circularity, solidity of MERS calculated by hand compared to DNN outputs. **(C1)** Statistical significances between the virus types, in terms of area and morphology, in the ground truth data (n=33, ^p<0.05 between groups). **(C2)** Statistical significances between the virus types, in terms of area and morphology, in the DNN output data (n=33, ^p<0.05 between groups). **(D1)** Statistical significances between THP1 M1 and THP1 M2, in terms of area and morphology, in the ground truth data (n=51 and n=60, respectively. ^p<0.05 between groups). **(D2)** Statistical significances between THP1 M1 and THP1 M2, in terms of area and morphology, in the DNN output data (n=51 and n=60, respectively. ^p<0.05 between groups).

## 4. Discussion

Here, we have demonstrated the ability to use multi-class instance segmentation to correctly analyze the morphological differences between multiple types of mammalian cells, as well as COVID-19 and MERS-CoV. By comparing precisions over recalls, classification accuracies, and Jaccard similarity coefficients, the DNN framework was able to produce a higher Jaccard similarity coefficient than of one using a Mask-R-CNN framework with ResNet101 backbone. This was achieved through DNN's decision-making algorithm, which tests out different networks and finds the best fit CNNs for one's specific task. We have found that CNNs with higher reported benchmarking accuracies [27] may not produce higher accuracies for certain biomedical engineering tasks. For example, both U-Net with a VGG16 backbone and U-Net with ResNet50 backbone yielded higher Jaccard similarity coefficients and global accuracies than U-Net with an Inception-ResNet-v2 backbone, despite Inception-ResNet-v2 having a higher reported benchmarking accuracy than both ResNet50 and VGG16. SqueezeNet, which has a lower benchmarking accuracy than GoogLeNet, was found more apt for classifying mammalian cells and was thus chosen to be in place of CNN I in DNN. As observed in **Figure 6 (C1-C2),** the DNN analysis showed statistical significance in area and circularity of the COVID19 in comparison to the MERS virus particles, which aligned with findings in the ground truth data of the viruses. In **Figure 6(D1-D2),** the DNN analysis also showed statistical significance in area and solidity of the THP1 M1 cells in comparison to the THP1 M2 cells; however, circularity was not statistically significant between the cells according to the DNN analysis. In terms of instance segmentation abilities, DNN's object detection network ability to cut out overlapping cells appeared to help the semantic segmentation network do a superior job of cell and virus edge detection. This resulted in DNN's higher Jaccard similarity coefficient compared to Mask-R-CNN's. As better CNNs are invented every day, the DNN can evolve to yield better accuracy over time by adding new state-of-the-art networks to the arena and culling older CNNs from CNN I, CNN II, or CNN III. Other CNNs can take the place of CNN I, CNN II, and CNN III to compete and ultimately yield a better final DNN for a given biomedical engineering task.

We have decided to identify the morphometric parameters that are considered important in viral pathogenesis: area, eccentricity, circularity, and solidity. These morphological parameters are important because differences in these aspects of virus morphology result in different pathogenic ability. The morphological parameters are also important for mammalian cells to study the effects of cell to cell interaction, virus-induced cell morphology, or stem cell morphology that indicates a certain type of differentiation. We have demonstrated that time and labor-consuming forms of cellular and viral morphometric analysis can be replaced by sub-section and one-click operation using DNN. Popular software tools, such as Cell Profiler[28] and Image J[29], require manual parameter tuning often requiring familiarity with the software and manual inputs from the user; however, this also has the potential to create user bias when analyzing morphometric data of cells. Robustly trained DNN, with large-scale datasets of cells, maybe a solution to a non-biased sub-second solution. This may also eliminate the need for chemical assays or FACs for cell analysis when used in conjunction with a benchtop microscope. Chemical assays and FACs are often time and labor intensive, and cell processing may change the morphometrics of the stained or sorted cells as well as result in the further production of chemical and biological waste; however, DNN, when used in conjunction with the microscope, eliminates the aforementioned downsides. In the future, the DNN framework can also be implemented to examine the morphological change of virus-infected cells. The DNN could also be used to examine the virus' response to therapeutic interventions, such as through examination of structural changes that may inhibit the virus' ability to infect the host.

In terms of computational power and time, DNN's architecture of partitioning the CNNs may also be advantageous compared to using one large instance segmentation CNN due to partitioning the GPU usage as well as easier optimization of each CNNs. The DNN generally requires a less GPU exhaustive CNN for object detection and then employs a more GPU exhaustive CNN for classification; for example, classifying and detecting COVID-19 virus particles from MERS virus particles may require more exhaustive CNN as a backbone, but one can use relatively less resource exhaustive CNN backbone to only locate the virus particles with high accuracy from the background of TEM images before feeding them into a more exhaustive classification network. By cropping the objects of interest and feeding it into the segmentation network, DNN was able to achieve a high score for the Jaccard similarity coefficient for multi-class instance segmentation. In the future, we will seek to have DNN analysis encompass a wider variety of viruses and cell types to broaden the application and ease the implementation of the DNN framework in future research.

For the next step, we will train the DNN using SARS-CoV-2 infected cells and mock cells observed in sputum sample smears of human subjects. This would enable us to rapidly diagnose SARS-CoV-2 infected patients using the DNN in conjunction with any benchtop microscopes. Cells in sputum samples of SARS infected patients showed cellular abnormalities, such as cytoplasmic foaminess, distinct vacuoles, multinucleation, and glass appearance of the nucleus. [30] SARS-CoV-2 infected cells also showed a dramatic increase in filopodial protrusions, which were significantly longer and more branched than in uninfected cells. [31] Uninfected cells also exhibited filopodial protrusions, but their frequency and shape were dramatically different. The SARS-CoV-2 infected cells also revealed prominent M protein clusters, possibly making assembled viral particles, localized along the tips of actin-rich filopodia. [31] Reorganization of the actin cytoskeleton is a common feature of many viral infections and is associated with different stages of the viral life cycle. [31, 32] We hypothesize that the cell morphology changes due to SARS-CoV-2 infection can be detected by the DNN. As a pre-trained DNN takes any time from subsecond to less than a minute, according to the user's computer hardware specifications, a pre-trained DNN using SARS-Cov-2 cells and mock cells from sputum sample smears of human subjects can be rapidly distributed around the world and used in conjunction with existing benchtop microscopes for rapid and scalable screening. Furthermore, different DNNs will be trained to classify SARS-Cov-2 infected cells and mock cells present in sputum samples according to patients' age group, sex, and ethnicity. This is to personalize the diagnostic method for higher accuracy in screening. Furthermore, classfication of cells infected by different types of coronaviruses and mock cells will studied using DNN.

# H2: Supplementary Materials

## Morphometric Analysis of Cells

### (A) Area

| Cell 1 | Cell 2 | Tukey HSD Q statistic | Tukey HSD p-value | Tukey HSD inference |
|---|---|---|---|---|
| RAW M1 | RAW M2 | 9.2524 | 0.0010053 | ** p<0.01 |
| RAW M1 | THP1 M1 | 21.7972 | 0.0010053 | ** p<0.01 |
| RAW M1 | THP1 M2 | 15.5203 | 0.0010053 | ** p<0.01 |
| RAW M1 | BMDM | 12.6017 | 0.0010053 | ** p<0.01 |
| RAW M1 | DF | 12.012 | 0.0010053 | ** p<0.01 |
| RAW M2 | THP1 M1 | 15.0891 | 0.0010053 | ** p<0.01 |
| RAW M2 | THP1 M2 | 8.5606 | 0.0010053 | ** p<0.01 |
| RAW M2 | BMDM | 0.8823 | 0.8999947 | insignificant |
| RAW M2 | DF | 6.8006 | 0.0010053 | ** p<0.01 |
| THP1 M1 | THP1 M2 | 5.9432 | 0.0010053 | ** p<0.01 |
| THP1 M1 | BMDM | 15.8912 | 0.0010053 | ** p<0.01 |
| THP1 M1 | DF | 4.8046 | 0.0092285 | ** p<0.01 |
| THP1 M2 | BMDM | 8.8492 | 0.0010053 | ** p<0.01 |
| THP1 M2 | DF | 0.1706 | 0.8999947 | insignificant |
| BMDM | DF | 6.7537 | 0.0010053 | ** p<0.01 |

### (B) Eccentricity

| Cell 1 | Cell 2 | Tukey HSD Q statistic | Tukey HSD p-value | Tukey HSD inference |
|---|---|---|---|---|
| RAW M1 | RAW M2 | 2.2644 | 0.584413 | insignificant |
| RAW M1 | THP1 M1 | 11.4419 | 0.0010053 | ** p<0.01 |
| RAW M1 | THP1 M2 | 8.0781 | 0.0010053 | ** p<0.01 |
| RAW M1 | BMDM | 2.0161 | 0.6856653 | insignificant |
| RAW M1 | DF | 5.997 | 0.0010053 | ** p<0.01 |
| RAW M2 | THP1 M1 | 12.5919 | 0.0010053 | ** p<0.01 |
| RAW M2 | THP1 M2 | 9.388 | 0.0010053 | ** p<0.01 |
| RAW M2 | BMDM | 4.4165 | 0.0226619 | * p<0.05 |
| RAW M2 | DF | 7.0972 | 0.0010053 | ** p<0.01 |
| THP1 M1 | THP1 M2 | 3.1728 | 0.2189204 | insignificant |
| THP1 M1 | BMDM | 10.9571 | 0.0010053 | ** p<0.01 |
| THP1 M1 | DF | 2.7829 | 0.3624971 | insignificant |
| THP1 M2 | BMDM | 7.3832 | 0.0010053 | ** p<0.01 |
| THP1 M2 | DF | 0.1333 | 0.8999947 | insignificant |
| BMDM | DF | 5.3065 | 0.0025532 | ** p<0.01 |

### (C) Circularity

| Cell 1 | Cell 2 | Tukey HSD Q statistic | Tukey HSD p-value | Tukey HSD inference |
|---|---|---|---|---|
| RAW M1 | RAW M2 | 10.5548 | 0.0010053 | ** p<0.01 |
| RAW M1 | THP1 M1 | 16.0478 | 0.0010053 | ** p<0.01 |
| RAW M1 | THP1 M2 | 12.8635 | 0.0010053 | ** p<0.01 |
| RAW M1 | BMDM | 8.8303 | 0.0010053 | ** p<0.01 |
| RAW M1 | DF | 12.3657 | 0.0010053 | ** p<0.01 |
| RAW M2 | THP1 M1 | 22.5009 | 0.0010053 | ** p<0.01 |
| RAW M2 | THP1 M2 | 19.7903 | 0.0010053 | ** p<0.01 |
| RAW M2 | BMDM | 20.0713 | 0.0010053 | ** p<0.01 |
| RAW M2 | DF | 17.7996 | 0.0010053 | ** p<0.01 |
| THP1 M1 | THP1 M2 | 3.2694 | 0.1900636 | insignificant |
| THP1 M1 | BMDM | 11.9541 | 0.0010053 | ** p<0.01 |
| THP1 M1 | DF | 0.5589 | 0.8999947 | insignificant |
| THP1 M2 | BMDM | 8.3212 | 0.0010053 | ** p<0.01 |
| THP1 M2 | DF | 2.2397 | 0.5944777 | insignificant |
| BMDM | DF | 8.8272 | 0.0010053 | ** p<0.01 |

### (D) Solidity

| Cell 1 | Cell 2 | Tukey HSD Q statistic | Tukey HSD p-value | Tukey HSD inference |
|---|---|---|---|---|
| RAW M1 | RAW M2 | 6.9065 | 0.0010053 | ** p<0.01 |
| RAW M1 | THP1 M1 | 18.9307 | 0.0010053 | ** p<0.01 |
| RAW M1 | THP1 M2 | 13.6544 | 0.0010053 | ** p<0.01 |
| RAW M1 | BMDM | 5.4864 | 0.0015576 | ** p<0.01 |
| RAW M1 | DF | 13.9086 | 0.0010053 | ** p<0.01 |
| RAW M2 | THP1 M1 | 22.9062 | 0.0010053 | ** p<0.01 |
| RAW M2 | THP1 M2 | 18.0135 | 0.0010053 | ** p<0.01 |
| RAW M2 | BMDM | 12.8686 | 0.0010053 | ** p<0.01 |
| RAW M2 | DF | 17.3501 | 0.0010053 | ** p<0.01 |
| THP1 M1 | THP1 M2 | 5.0268 | 0.0053124 | ** p<0.01 |
| THP1 M1 | BMDM | 16.9054 | 0.0010053 | ** p<0.01 |
| THP1 M1 | DF | 1.2332 | 0.8999947 | insignificant |
| THP1 M2 | BMDM | 11.209 | 0.0010053 | ** p<0.01 |
| THP1 M2 | DF | 3.0588 | 0.2562279 | insignificant |
| BMDM | DF | 11.9403 | 0.0010053 | ** p<0.01 |

**Table S1. Cells' Morphometrics Observed by the DNN.** The table shows Tukey HSD Q statistic, HSD p-value, and Tukey HSD inference between the cell types in terms of **(A)** area, **(B)** eccentricity, **(C)** circularity, and **(D)** solidity.

**Acknowledgments**

**General**: We thank Dr. Shih-Fu Chang, Senior Executive Vice Dean and the Richard Dicker Professor of The Fu Foundation School of Engineering and Applied Science at Columbia University, Dr. David Ho, founding scientific director of the Aaron Diamond AIDS Research Center and the Clyde and Helen Wu Professor of Medicine at Columbia University Vagelos College of Physicians and Surgeons, and Dr. Sam Sia, professor of Biomedical Engineering at Columbia University for useful conversations. We also thank Mariel Werner for assistance in THP-1 studies, and Dr. Michael Sutton for assistance in RAW cell studies.

**Funding:** This study was supported by NIH (R01-AR073529, HHL), DoD (W81XWH-18-1-0241, HHL), and FF-SEAS.

**Author contributions:** H.L. and S.L. designed the study. H.L., S.L., J.G., and A.L. designed, advised, and modified the DNN architecture. S.L. and Y.C. coded the DNN. S.L. and Y.C. cultured and imaged dermal fibroblasts and bone marrow derived macrophages. S.L. and Y.C. coded and tested all the CNNs used in this study. S.L., Y.C., A.B., and J.W. prepared and preprocessed the training, validation, and test set data. S.L. and A.J. tested and evaluated the Mask-R-CNN. S.L., Y.C., and A.B., wrote the manuscript. P.B. and D.B. cultured, polarized, and imaged THP-1 cells. P.B. cultured, polarized, and imaged RAW 264.7 cells. All authors edited the manuscript.

**Competing interests:** H.L., S.L., Y.C., J.G., A.L., are inventors on a pending provisional patent application submitted by the Columbia University related to this work. The authors declare that they have no other competing interests.

**Data and materials availability:** Additional data related to this paper may be requested form the authors.